\documentstyle[12pt]{article}

\def\qq{q\bar{q}}

\def\gp{\gamma p}

\def\gp{{\gamma p}}
\def\ylab{y_{\rm lab}}
\begin{document}
\begin{titlepage}
\vspace*{-1cm}
\begin{flushright}
 DTP/93/44   \\
 UDWPHYS-93-02\\
 June 1993 \\
\end{flushright}
\vskip 1.cm
\begin{center}
{\Large\bf Photoproduction of large-mass lepton pairs
at HERA as a probe of the
small $x$ structure of the proton}
\vskip 1.cm
{\large A.C. Bawa}
\vskip .2cm
{\it Department of  Physics, University of Durban-Westville \\
Durban, 4000 South Africa }\\
\vskip .4cm
{\large K. Charchu\l a\footnote{Permanent address: Physics Dept.,
Warsaw University, Hoza~69, PL-00-681 Warsaw, Poland}}
\vskip .2cm
{\it Department of Physics, University of Durham \\
Durham DH1 3LE, England }\\
\vskip .4cm
and
\vskip   .4cm
{\large  W.J. Stirling}
\vskip .2cm
{\it Departments of Physics and Mathematical Sciences, University of Durham \\
Durham DH1 3LE, England }\\
\vskip 1cm
\end{center}
\begin{abstract}
The photoproduction of Drell-Yan lepton pairs
at HERA is studied. We show that in the backward rapidity region
the cross section is strongly
sensitive to the small-$x$ behaviour of the quark distributions in the
proton, and that with sufficient luminosity it should be possible to
distinguish singular $xq \sim x^{-1/2}$ behaviour  from standard
Regge $xq \sim x^0$  behaviour. The effect on the event rate of
experimental acceptance cuts is studied with the help of a Monte Carlo
event generator.
\end{abstract}
\vfill
\end{titlepage}
\newpage

\section{Introduction}
The Drell-Yan process in hadron-hadron collisions has for many
years yielded important information on the quark distributions of
hadrons \cite{DRELL}. In proton-nucleon collisions, for example, the sea quark
distribution at medium and large $x$  -- which is overwhelmed in deep
inelastic scattering by the valence quarks --  can be directly measured.
It was pointed out several years ago \cite{KUNSZT} that the Drell-Yan
process at HERA could also be of interest. In particular, since the
photoproduction of lepton pairs arises mainly from  $\qq$ annihilation,
the quark distributions in the photon can be studied.  For example,
consider the cross section differential in both
lepton-pair mass $M$ and rapidity $y$, with the proton direction
defining $y>0$. In the centre-of mass frame of
the $\gamma p$ collision   in leading order, the momentum fractions of
the annihilating quarks are given simply
by $x_{\gamma} = \sqrt{\tau} \exp(-y)$, $x_{p} = \sqrt{\tau} \exp(y)$, where
$\tau = M^2/s_\gp$.
At large negative rapidities, therefore, the quark distribution in the photon
is probed at large $x$, while that in the proton is probed at small $x$.
Now at HERA, the fact that the photon-proton system has a large
positive boost in the
lab frame means that $\ylab < 0$ for the lepton pairs already implies
a very large rapidity in the photon-proton centre-of-mass frame. In fact, as we
shall demonstrate below, lepton pairs with $\ylab \sim -1$ probe
$x_\gamma \sim 1$ and $x_p \sim 10^{-4} - 10^{-3}$. In addition, such lepton
pairs should be relatively easy to identify, being produced in the
\lq backward' part of the
detector which is relatively free from hadronic activity.

The importance of such a measurement is  that it allows
an independent  probe of the small $x$ structure of the proton. Recent deep
inelastic structure function measurements from the H1 \cite{H1} and
ZEUS \cite{ZEUS} collaborations at HERA indicate
that $F_2(x,Q^2)$ rises
steeply at small $x$ and modest $Q^2$, in line with predictions
based on the \lq\lq perturbative pomeron" of QCD \cite{SMALLX}.
It is vitally important to obtain additional evidence of this effect,
and the Drell-Yan cross section at negative $\ylab$ provides just such
a measurement. We note that the $\gp$ Drell-Yan cross section has already
been advocated as a probe of the small $x$ structure of the {\it photon}
\cite{KUNSZT,GLUCKA,GLUCKB},
but this corresponds to large, positive $\ylab$ values. In practice, the
hadronic activity in this region presumably makes a clean measurement
rather difficult.

In this paper we  study the $\gp$ Drell-Yan cross section for $\ylab < 0$.
To test the discriminating power in the  $x_p \ll 1$ region, we use the recent
MRS-D$_0'$ and MRS-D$_-'$ parton distributions \cite{MRS92P}.
The former have \lq\lq conventional" Regge small $x$ behaviour,
$xq, xg \sim x^0$ as $x \to 0$, while the latter have more singular
$xq, xg \sim x^{-1/2}$ behaviour as expected  from perturbative
pomeron arguments \cite{MRS92P}.
 Note that the recent $F_2$ data from H1 \cite{H1} and ZEUS
\cite{ZEUS} at HERA strongly favour the D$_-'$ distributions.
A crucial part of our argument is that for $\ylab < 0$
 the parton distributions of the photon are
 being probed in a $x$ region where they are
 already constrained by $F_2^\gamma$ data. The difficulty here
stems from the  fact that in practice the $F_2^\gamma$ data in the relevant
$x,Q^2$ regions are not very precise, and this introduces some uncertainty
into the  Drell-Yan predictions which in turn weakens the sensitivity
to the proton structure. To try to quantify this, we use a variety
of recent photon parton-distribution parametrizations.

The paper is organised as follows. In the next section we describe the
theoretical formalism
for computing the Drell-Yan cross section at next-to-leading
order in photon-proton collisions.
Numerical calculations then quantify the overall event rates and
the differences obtained from using different sets of
parton distributions.
In Section 3 we perform a Monte Carlo event simulation to study in detail
the $x$ values which are being probed, and also to investigate the
energy and angular distributions of the leptons in the lab.
Section 4 contains our conclusions.

\section{The Drell-Yan cross section at next-to-leading order}

\subsection{Expression for the inclusive cross section}
In this study we  use both next-to-leading order matrix elements and
parton distributions, in the $\overline{\rm MS}$ scheme. We  calculate
 the inclusive cross sections of dilepton production at HERA:
\begin{eqnarray}
\gamma p & \rightarrow & l^+l^-+X\\
e p & \rightarrow & l^+l^-+X      \; ,
\end{eqnarray}
using  the expression  from \cite{GLUCKA}:
\begin{eqnarray}
\lefteqn{
{d\sigma\over{dM^2}} = {4\pi\alpha^2\over{3M^2}}\int_\tau^1{dx_p\over{x_p}}
\int_{\tau/x_p}^1{dx_\gamma\over{x_\gamma}}\sum_{q=u,d,s,c}e_q^2
         }
\nonumber\\
&&
\times \Biggl\{
\left[q_p(x_p,Q^2)\bar q_\gamma (x_\gamma,Q^2)+\bar q_p(x_p,Q^2)
q_\gamma(x_\gamma,Q^2)\right] \left[\delta(1-{\tau\over{x_px_\gamma}})
+{\alpha_s(\mu^2)\over{2\pi}}f_{\qq} \left({\tau\over{x_px_\gamma}}\right)
\right]
 \nonumber\\
&&
 +{\alpha_s(\mu^2)\over{2\pi}}\;   f_{gq}\left({\tau\over{x_px_\gamma}}\right)
\nonumber\\
&&
 \times\left[g_p(x_p,Q^2)
\left(q_\gamma(x_\gamma,Q^2)+\bar q_\gamma (x_\gamma,Q^2)\right)
+\left(q_p(x_p,Q^2)+\bar q_p(x_p,Q^2)\right)g_\gamma(x_\gamma,Q^2)\right]
\nonumber \\
&&
 +{6e_q^2\alpha\over{2\pi}}\; f_{gq}\left({\tau\over{x_px_\gamma}}\right)
\left(q_p(x_p,Q^2)+\bar q_p(x_p,Q^2)\right)\delta(1-x_\gamma)
\Biggr\}
\end{eqnarray}
where $\tau=M^2/s$ and the higher-order terms $f_{\qq}$ and $f_{gq}$ are given
in Eqn.(5) in \cite{GLUCKA}. In this expression, the first term contains
the dominating LO $\qq$ initial state and the associated
real  and virtual corrections.
 The second term contains the $g_p+q_\gamma$ and $q_p+g_\gamma$
initial states and finally the third term is the Compton term given by
$q_p+\gamma\rightarrow \gamma*+q$. Throughout this calculation we set the
factorization and renormalization scales at the invariant mass of the dilepton
pair; that is, $\mu^2=Q^2=M^2$.

For the parton distributions in the photon  which appear
in the above expression, we take  the GRV(NLO)
next-to-leading order set  \cite{GRV}  as  standard
and probe the sensitivity of the
cross sections to the MRS-D$_0'$ and MRS-D$_-'$ parton distributions.
We also study the sensitivity of the cross sections to different photon
parametrizations, to attempt to quantify the uncertainty in extracting
information on the small-$x$ proton distributions from this source.
To obtain the relevant cross sections for the $ep$ initial state we
convolute the cross-section expression with the simplest form of the
equivalent photon approximation \cite{WWA}, in which we have set the
energy scale to be $\hat s=x_px_es$.

\subsection{Event rates and sensitivity to photon and proton
parton distributions}
The cross-section calculations are performed for the energy parameters of
HERA. For the $\gamma p$ initial state we take $\sqrt s_{\gamma p}
=200$ GeV, and for the $ep$ initial state we assume $E_e=30$ GeV and $E_p=820$
GeV.
In Figure 1 we present the results of the calculation of the cross section
$d\sigma/dM$ as a function of $M$, where $M$ is the invariant mass of the
dilepton pair. The $\gamma p$ cross section is presented in Figure 1(a) for
both the MRS-D$_0'$ and MRS-D$_-'$ parton distributions. The MRS-D$_-'$ cross
section is larger, as expected,
 but the difference between the two decreases substantially
as $M$ increases, since at small $M$ the sensitivity to small-$x$ in the proton
is large and the differences between the two sets of proton distribution
functions is accentuated. At $M=2$ GeV/c$^2$ the cross section is approximately
76~pb/GeV for the MRS-D$_0'$ case and 155~pb/GeV
 for the MRS-D$_-'$ case, and
at $M=4$ GeV/c$^2$ the cross sections have fallen to  10~pb/GeV and
14~pb/GeV, respectively. The
next-to-leading contribution to this cross section is approximately 1.5 to 2
times that of the leading order one. In Figure 1(b) the corresponding
cross section for the $ep$ initial state is shown. Evidently
these cross sections are approximately one order of magnitude smaller
than those for $\gamma p$. The reason  is that
at these energies  ${\cal L}_{\gamma p}
\simeq 0.1\! {\cal L}_{ep}$. The sensitivity of the $ep$ cross sections to the
two sets of parametrizations is somewhat smaller than in the $\gamma p$ case.

To emphasize the sensitivity of the Drell-Yan cross sections to the small-$x$
behaviour of the proton distribution functions we next calculate the rapidity
distribution of the dilepton pairs in the $\gamma p$ case. The
kinematics are such that $x_p=\sqrt\tau {\rm e}^{y_{\gp}}$ and
$x_\gamma=\sqrt\tau {\rm e}^{-y_{\gp}}$,
which implies that $x_p\rightarrow 0$ in the backward direction, which is the
region
where differences between the MRS-D$_0'$ and MRS-D$_-'$ should be
observed.
In Figure 2 we present the plots for the cross section $d\sigma/dMdy$ as a
function of $y$ for $M=4,6$ GeV/c$^2$. The $M=4$ GeV/c$^2$ case is shown
in Figure 2(a) and we see that at large negative $ y$  ($\equiv\ylab$),
 the MRS-D$_-'$
cross section is larger than the MRS-D$_0'$ one by a factor larger than 2.
At this invariant mass the proton is probed down to $x_p\simeq 4\times10^{-4}$
(see below).
At smaller $M$ the difference is considerably larger; for instance at $M=2$
GeV/c$^2$ there is a factor of 4 between the two distributions at large
negative rapidities. However, at such low masses data from
hadron-hadron collisions suggest that  the background to the dilepton
cross section arising from
$J/\psi$ production may be substantial \cite{DRELL}. In what follows,
therefore, we impose a lower limit of $M > 4$ GeV/c$^2$ on the dilepton
mass.
In Figure 2(b) the same distribution is plotted for
$M=6$ GeV/c$^2$ and as expected, the effect washes out, due to the fact
that a larger invariant mass forces larger $x_p$ values; at large
negative rapidities the cross section ratio,
MRS-D$_-'$/MRS-D$_0'$ reduces to approximately 1.5. In the positive rapidity
region the ratio is very nearly 1 in both Figures 2(a),(b).

The usefulness of this process to probe the small-$x$ behaviour of the proton
structure function depends on the
insensitivity of the cross sections to the large-$x$ behaviour of the
photon structure function. To investigate this, we repeat the
calculations using several other sets of photon distributions,
keeping the proton distributions (MRS-D$_-'$) fixed.
Figure 3 shows  the rapidity
distribution, $d\sigma/dMdy$ for $M=4$ GeV/c$^2$ as a function of $ y$
 for various
choices of the photon distributions: (i) the reference GRV(NLO) set
\cite{GRV} (solid line), (ii) the GS(NLO) set \cite{GS} (dashed line)
and the LAC1(LO) set \cite{LAC} (dotted line).
In the interesting region of phase space (large negative
rapidities) we see that  the different distributions give roughly the
same shape in rapidity but a spread in normalisation of order $10\%$.
This can be understood in terms of the spread in predictions of these
sets for the photon structure function $F_2^\gamma$. Figure 4 shows
 data on $F_2^\gamma$ together with the predictions of the same three
sets as in Figure 3. We have chosen data in a similar $Q^2$ range
as the $M^2$ values relevant to the Drell-Yan cross section at
HERA. We see immediately that the spread in the predictions
simply reflects the uncertainty in the structure function
measurements. More precise measurements of $F_2^\gamma$ would of course
pin down the Drell-Yan cross section more accurately. In the meantime
we should regard the $O(10\%)$ spread in normalisation as
a systematic uncertainty in the extraction of the {\it proton}
distributions from the Drell-Yan data.

\section{Event simulation}
In the previous section we have seen how the $\gamma p$ Drell-Yan
cross section for $\ylab < 0$ and $ M > 4$ GeV/c$^2$ offers a good
possibility of obtaining information on the quark distribution
in the proton at small $x$. An important question is whether the bulk
of these events actually fall within the acceptance of the HERA
detectors. To study this, we perform a Monte Carlo simulation using the
PYTHIA5.6 \cite{PYTHIA} MC generator for $\gp \to l^+l^- + X$
at $\sqrt{s_{\gp}} = 200$ GeV.
This simulation also allows us to investigate which  $x_p$, $x_\gamma$
values give the dominant contribution to the cross section.

Figure 5 show the distributions in (a) $x_{\gamma}$ and (b) $x_p$
for $M > 4$ GeV/c$^2$. The two histograms correspond to $\ylab < 0$
(solid lines) and $\ylab > 0$ (dashed lines). The former is dominated
by $x_p$ values in the $10^{-4} - 10^{-3}$ region, as expected.
Selecting only those events with $M > 4$ GeV/c$^2$, and
defining $\vartheta_l$ and $E_l$ to be the lepton polar angle and energy
in the lab frame respectively, Figure 6 shows the scatter plots for
(a) $x_\gamma$ {\it vs.} $\vartheta_l$,
(b) $x_p$ {\it vs.} $\vartheta_l$,
(c) $E_l$ {\it vs.} $\vartheta_l$,
and (d) $M$ {\it vs.} $\vartheta_l$.
We see immediately that most of the events have leptons which
are produced with reasonable  energy and at sizeable angles to the
beam direction $\vartheta = 0^{\rm o}$, $180^{\rm o}$. Given that
we expect very little additional  hadronic activity in these regions
of phase space, the acceptance for both muon and electron
pairs should be quite high.

\section{Conclusions}

The structure of the proton at small $x$ is an important new area
of interest, and  the first $F_2$ data from HERA provide tentative
evidence for the \lq\lq perturbative pomeron". The Drell-Yan cross
section in the backward rapidity region should in the very near future
 give a confirmation of this small-$x$
 behaviour from a completely different process.
The decrease in hadronic activity as $\ylab$
becomes more negative and the effect of the boost on the lepton lab angle
should make  this a particularly clean measurement.
Our study has concentrated  on $\gp$ collisions at $\sqrt{s_{\gp}} =
200$ GeV, tacitly assuming that the final-state electron can be tagged
with reasonable efficiency at HERA. However, as is clear from Figure 1(b),
the effects we describe should also be observable in the $ep$ cross section.

Observation  of the dramatic differences induced by \lq\lq standard"
and \lq\lq singular"  small-$x$ behaviour  in the proton in the $\ylab < 0$
Drell-Yan cross section relies
on  knowledge of the large-$x$ photon  structure.
 We have shown that
the uncertainty in the photon  at large $x$ is of the order of $10$\%,
which for $M=4 - 6$ GeV is small in comparison to the D$_0'$/D$_-'$ effect.
But we should
stress again that once these $F_2^p$ become well-measured at HERA, the
Drell-Yan
cross section will provide a complementary measurement of the
large-$x$ behaviour of the photon.

\bigskip

\vspace{1cm}
\noindent {\Large\bf Acknowledgments}

\bigskip
KCh would like to thank the Physics Department and Grey College
at the  University  of Durham for their warm hospitality.
The work of KCh was supported   by a EC \lq\lq Go--West" Scholarship.
ACB would like to thank the Physics Department at the University
of Durham for its kind hospitality when this project was conceived,
and the Foundation for Research Development for partial funding.
We are grateful to Lionel Gordon for discussions concerning the
GS parametrizations.


\newpage

\noindent{\Large\bf Figure Captions}

\begin{itemize}
\item[{[1]}]
Predictions for the
Drell-Yan cross section $d\sigma / dM$ in (a) $\gp$
collisions at $\sqrt{s_{\gp}} = 200$ GeV, and (b) $ep$ collisions
at $\sqrt{s_{ep}} = 314$ GeV. The curves
correspond to next-to-leading order calculations using the GRV(NLO)
distributions for the photon \cite{GRV} with  MRS-D$_0'$ (dashed line)
and MRS-D$_-'$ (solid line)  distributions for the proton \cite{MRS92P}.
\item[{[2]}]
Predictions for the Drell-Yan rapidity distribution $d\sigma / dM dy $  for
(a) $M = 4$ GeV/c$^2$ and
(b) $M = 6$ GeV/c$^2$ in  $\gp$
collisions at $\sqrt{s_{\gp}} = 200$ GeV.
The curves correspond to next-to-leading order calculations using the GRV(NLO)
distributions for the photon \cite{GRV} with  MRS-D$_0'$ (dashed line)
and MRS-D$_-'$ (solid line)  distributions for the proton \cite{MRS92P}.
\item[{[3]}]
Predictions for the Drell-Yan rapidity distribution $d\sigma / dM dy $  for
 $M =4$ GeV/c$^2$ in  $\gp$
collisions at $\sqrt{s_{\gp}} = 200$ GeV.
The curves correspond to next-to-leading order calculations using
(a)  GRV(NLO) \cite{GRV} (solid line),
(b)  GS(NLO) \cite{GS} (dashed line)
and (c)  LAC1(LO) \cite{LAC} (dotted line)
distributions for the photon,
with MRS-D$_-'$ distributions for the proton \cite{MRS92P}.
\item[{[4]}]
The photon structure function $F_2^{\gamma}(x,Q^2)$. The data are from
the PLUTO \cite{PLUTO} and TASSO \cite{TASSO}
collaborations. The curves are the theoretical predictions
using the same sets as in Figure 3:
(a)  GRV(NLO) \cite{GRV} (solid line),
(b)  GS(NLO) \cite{GS} (dashed line)
and (c)  LAC1(LO) \cite{LAC} (dotted line).
\item[{[5]}]
Distributions in (a) $x_{\gamma}$ and (b) $x_p$
for $M > 4$ GeV/c$^2$, using a  PYTHIA5.6 \cite{PYTHIA}
MC simulation of $\gp\to l^+l^- +X$ at
$\sqrt{s_{\gp}} = 200$ GeV. The two histograms correspond to $\ylab < 0$
(solid lines) and $\ylab > 0$ (dashed lines).
\item[{[6]}]
Scatter plots for
(a) $x_\gamma$ {\it vs.} $\vartheta_l$,
(b) $x_p$ {\it vs.} $\vartheta_l$,
(c) $E_l$ {\it vs.} $\vartheta_l$,
and (d) $M$ {\it vs.} $\vartheta_l$,
for $M > 4$ GeV/c$^2$, using a PYTHIA5.6 \cite{PYTHIA}
MC simulation of $\gp\to l^+l^- +X$ at
$\sqrt{s_{\gp}} = 200$ GeV.

\end{itemize}


\newpage

\begin{thebibliography}{99}

\bibitem{DRELL} For a recent review of Drell-Yan physics see
for example M.R.~Whalley and W.J.~Stirling, Durham-RAL HEP Database publication
DPDG/93/01, to be published in J. Phys. {\bf G} (1993).

\bibitem{KUNSZT} Z. Kunszt and W.J. Stirling, Phys. Lett. {\bf 217B}
(1989) 563.

\bibitem{H1} H1 Collaboration: A. De Roeck, Proc. of the Durham Workshop
on \lq HERA--the new frontier for QCD' (March 1993), to be published in
J. Phys. {\bf G} (1993).

\bibitem{ZEUS} ZEUS Collaboration: A. Caldwell,
  talk presented at DESY, May 18th 1993.

\bibitem{SMALLX} For a recent review see J. Kwiecinski,
Proc. of the Durham Workshop
on \lq HERA--the new frontier for QCD' (March 1993), to be published in
J. Phys. {\bf G} (1993).

\bibitem{GLUCKA} M. Gl\"uck, E. Reya and A. Vogt, Phys. Lett. {\bf 285B}
(1992) 285.

\bibitem{GLUCKB} M. Gl\"uck, E. Reya and A. Weber, Phys. Lett. {\bf 298B}
(1993) 176.

\bibitem{MRS92P}A.D.\ Martin, W.J.\ Stirling and R.G.\ Roberts,
Phys.\ Lett. {\bf 306B} (1993) 145.

\bibitem{GRV} M. Gl\"uck, E. Reya and A. Vogt, Phys. Rev. {\bf D46}
(1992) 1973.

\bibitem{WWA} C.F. Weizs\"acker, Z. Phys. {\bf 88} (1934) 612, E.J. Williams,
Phys. Rev. {\bf 45} (1934) 729.




\bibitem{GS} L.E. Gordon and J.K. Storrow, Z. Phys. {\bf C56} (1992) 307.

\bibitem{LAC} H. Abramowicz, K. Charchula and A. Levy, Phys. Lett. {\bf 269B}
(1991) 458.


\bibitem{PLUTO} PLUTO Collaboration: Phys. Lett. {\bf 142B} (1984) 111,
                Nucl. Phys. {\bf B281} (1987) 365.

\bibitem{TASSO} TASSO Collaboration: Z. Phys. {\bf C31} (1986) 527.


\bibitem{PYTHIA} T. Sj\"ostrand, CERN-TH.6488/92 (1992);
 Comp. Phys. Comm. {\bf 46} (1987) 43.


\end{thebibliography}
\end{document}